\documentclass[twocolumn]{aastex61}

\newcommand{\sersic}{S\'ersic }

\received{\today}
\revised{\today}
\accepted{\today}
\submitjournal{ApJL}

\shorttitle{}
\shortauthors{J.M.Cann et al}

\begin{document}

\title{Relics of Supermassive Black Hole Seeds: The Discovery of an Accreting Black Hole in an Optically Normal, Low Metallicity Dwarf Galaxy}

\correspondingauthor{Jenna M. Cann}
\email{jcann@masonlive.gmu.edu}

\author{Jenna M. Cann}
\affiliation{George Mason University, Department of Physics and Astronomy, MS3F3, 4400 University Drive, Fairfax, VA 22030, USA}
\affiliation{National Science Foundation, Graduate Research Fellow}

\author{Shobita Satyapal}
\affiliation{George Mason University, Department of Physics and Astronomy, MS3F3, 4400 University Drive, Fairfax, VA 22030, USA}

\author{Barry Rothberg}
\affiliation{George Mason University, Department of Physics and Astronomy, MS3F3, 4400 University Drive, Fairfax, VA 22030, USA}
\affiliation{LBT Observatory, University of Arizona, 933 N. Cherry Ave., Tuscan, AZ 85721, USA}

\author{Gabriela Canalizo}
\affiliation{Department of Physics and Astronomy, University of California, Riverside, 900 University Avenue, Riverside, CA 92521, USA}

\author{Thomas Bohn}
\affiliation{Department of Physics and Astronomy, University of California, Riverside, 900 University Avenue, Riverside, CA 92521, USA}

\author{Stephanie LaMassa}
\affiliation{Space Telescope Science Institute, 3700 San Martin Drive, Baltimore, MD 21218, USA}

\author{William Matzko}
\affiliation{George Mason University, Department of Physics and Astronomy, MS3F3, 4400 University Drive, Fairfax, VA 22030, USA}

\author{Laura Blecha}
\affiliation{University of Florida, Department of Physics, P.O. Box 118440, Gainesville, FL 32611-8440}

\author{Nathan J. Secrest}
\affiliation{U.S. Naval Observatory, 3450 Massachusetts Avenue NW, Washington, DC 20392, USA}

\author{Anil Seth}
\affiliation{Department of Physics and Astronomy, University of Utah, 115 South 1400 East, Salt Lake City, UT 84112, USA}

\author{Torsten B\"oker}
\affiliation{European Space Agency, c/o STSCI, 3700 San Martin Drive, Baltimore, MD 21218, USA}

\author{Remington O. Sexton}
\affiliation{U.S. Naval Observatory, 3450 Massachusetts Avenue NW, Washington, DC 20392, USA}
\affiliation{George Mason University, Department of Physics and Astronomy, MS3F3, 4400 University Drive, Fairfax, VA 22030, USA}

\author{Lara Kamal}
\affiliation{George Mason University, Department of Physics and Astronomy, MS3F3, 4400 University Drive, Fairfax, VA 22030, USA}

\author{Henrique Schmitt}
\affiliation{Naval Research Lab, Washington, DC 20375, USA}



\begin{abstract}
The detection and characterization of supermassive black holes (SMBHs) in local low mass galaxies is crucial to our understanding of the origins of SMBHs. This statement assumes that low mass galaxies have had a relatively quiet cosmic history, so that their black holes have not undergone significant growth and therefore can be treated as relics of the original SMBH seeds. While recent studies have found optical signatures of active galactic nuclei (AGNs) in a growing population of dwarf galaxies, these studies are biased against low metallicity and relatively merger-free galaxies, thus missing precisely the demographic in which to search for the relics of SMBH seeds.  Here, we report the detection of the [\ion{Si}{6}]1.963~$\mu$m coronal line (CL), a robust indicator of an AGN in the galaxy SDSS~J160135.95+311353.7, a nearby ($z=0.031$) low metallicity galaxy with a stellar mass approximately an order of magnitude lower than the LMC ($M_*\approx10^{8.56}$~M$_\odot$) and no optical evidence for an AGN. The AGN bolometric luminosity implied by the CL detection is $\approx10^{42}$~erg~s$^{-1}$, precisely what is predicted from its near-infrared continuum emission based on well-studied AGNs. Our results are consistent with a black hole of mass $\approx~10^5$~M$_\odot$, in line with expectations based on its stellar mass. This is the first time a near-infrared CL has been detected in a low mass, low metallicity galaxy with no optical evidence for AGN activity, providing confirmation of the utility of infrared CLs in finding AGNs in low mass galaxies when optical diagnostics fail. These observations highlight a powerful avenue of investigation to hunt for low mass black holes in the JWST era.
\end{abstract}




\section{Introduction} \label{sec:intro}

Understanding the occupation fraction and properties of supermassive black holes (SMBHs) in low mass galaxies places important constraints on seed SMBHs, formed at high redshift and observationally inaccessible \citep[see review by][and references therein]{greene2020}.  For this reason, the search for SMBHs in dwarf galaxies has been an active area of research over the past decade. Uncovering the lowest masses dynamically is extremely challenging due to their small spheres of influence, although recent efforts have made notable progress \citep[e.g.][]{nguyen2019}.  Therefore, most efforts have focused on finding accreting SMBHs (active galactic nuclei; AGNs) in the lowest mass galaxies.  While impressive progress has been made in uncovering AGNs in dwarf galaxies have been found using a variety of multi-wavelength observations \citep[see review by][and references therein]{greene2020}, they still constitute a very small fraction of the dwarf galaxy population.  For example, only 0.1\% of dwarf galaxies with masses $<10^{9.5}$~M$_\odot$, approximately the mass of the Large Magellanic Cloud (LMC), have unambiguous signatures of AGNs \citep{reines2013} through the commonly used optical spectroscopic Baldwin-Phillips-Terlevich (BPT) diagram \citep{baldwin1981,kewley2001,kauffmann2003}.
While this could be cited as potential evidence for a dearth of IMBHs in the local universe, recent theoretical work has shown that standard optical BPT diagrams are ineffective for low mass AGNs, since the hardening of the spectral energy distribution changes the ionization structure of the nebula \citep{cann2019}.  Furthermore, the majority of AGNs in dwarf galaxies discovered so far have been found in higher metallicity galaxies, since low metallicity galaxies of all masses tend to occupy the star-forming region of the BPT diagram \citep{groves2006,cann2019}. Indeed, a detailed study of the [N~II]/H$\alpha$ values, a strong indicator of the metallicity of a galaxy, indicates that nearly all AGNs have super-Solar metallicities, with only 0.2\% of a sample of Seyfert 2s from SDSS identified as low metallicity \citep{groves2006}. 
In addition, since AGN identification using optical narrow line diagnostics requires that the AGN dominate over star formation, this selection mechanism favors AGNs in galaxies with less active star formation, which are often more bulge-dominated.  However, recent HST studies have shown a wide range of morphologies in BPT-selected AGNs in dwarf galaxies \citep{kimbrell2021}.

This is a limitation, as the premise behind using dwarf galaxies to search for local analogs of SMBH seeds rests on the assumption that the galaxy has had a relatively quiescent evolution to ensure that their black holes can be considered relics of the original seeds.  A dwarf galaxy with a high \sersic index suggests external interactions, such as merging or tidal stirring, or significant secular evolution, that could drive gas towards the center of the galaxy, fueling star formation, thereby enriching the gas phase metallicity, and potentially growing the black hole.  Therefore, many of the currently widely used methods employed in the hunt for AGNs in dwarf galaxies are biased towards a host galaxy demographic that is not ideal in order to gain insight into the initial SMBH seed population.

The use of infrared coronal lines (CLs), with ionization potentials $>70$ eV, has been proposed as an alternative diagnostic \citep{cann2018,cann2020,satyapal2021} to find SMBHs hidden by other diagnostics, as these lines are not susceptible to dilution from star formation, a common problem for the BPT diagnostics, which rely on emission lines corresponding to ionization potentials less than $35$~eV.  As stars do not produce enough high energy radiation to excite these ions, the detection of a coronal line can be unambiguous proof of the presence of an AGN.  Further, infrared CLs, such as [\ion{Ne}{5}], [\ion{Si}{6}], and [\ion{Si}{10}], are less susceptible to dust extinction.  They have been detected in a large sample of AGNs in high mass galaxies \citep[e.g.,][]{riffel2006,veilleux2009,rodriguezardila2011,lamperti2017,mullersanchez2018}, and have revealed AGNs in bulgeless galaxies with no optical signatures of an AGN \citep{satyapal2007,satyapal2008,satyapal2009} and low metallicity galaxies \citep{cann2020}.  With the advent of the James Webb Space Telescope (JWST), high sensitivity infrared spectra of large samples of dwarf galaxies, extending to the lowest masses, will be possible, enabling the discovery of the most extreme populations well into the intermediate mass black hole (IMBH; $10^2-10^4$~M$_\odot$).  In this Letter, we present the first near-infrared coronal line evidence of an AGN in a low metallicity dwarf galaxy with no optical evidence for an AGN.

\section{Observations and Data} \label{sec:observations}

\subsection{Target Selection}

\begin{figure}
\includegraphics[width=0.495\textwidth]{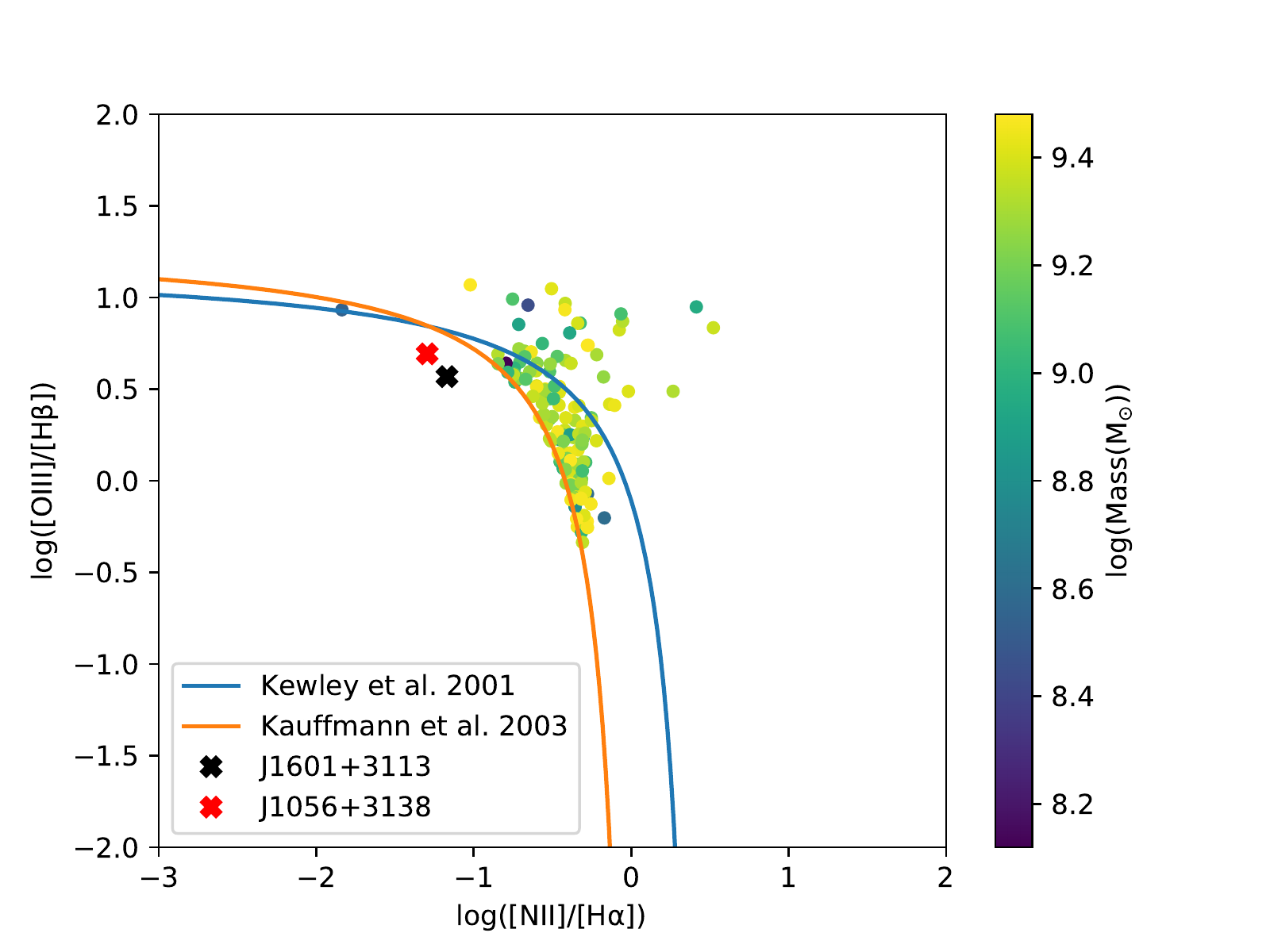}
\caption{Optical emission line ratios for J1601+3113 (black `x') place it securely in the star-forming region of the BPT diagram. Also pictured for comparison is data from the \citet{reines2013} sample, with galaxy stellar mass in the color bar.}
\label{bpt}
\end{figure}

J160135.95+311353.7 (hereafter J1601+3113) is a nearby ($z=0.031$) galaxy, with no broad lines, and optically normal line ratios in the star-forming region of the BPT diagram (Figure \ref{bpt}). Based on the Max Planck Institute f\"ur Astrophysik/Johns Hopkins University (MPA/JHU) catalog\footnote{http://www.mpa-garching.mpg.de/SDSS/} of derived properties for the Sloan Digital Sky Survey (SDSS) data release 8, it has a galaxy mass of $10^{8.56}$~M$_\odot$, roughly an order of magnitude lower mass than the LMC.  The NASA-Sloan Atlas catalog lists a \sersic index of $1.87$, implying a disk-like morphology, as expected based on its low metallicity \citep{izotov2012}.  While a detailed analysis of J1601+3113's metallicity is given in section 3.5, its log([\ion{N}{2}]/H$\alpha$) of -1.17 suggests a sub-Solar metallicity \citep{groves2006}.  

\begin{figure}
\includegraphics[width=0.495\textwidth]{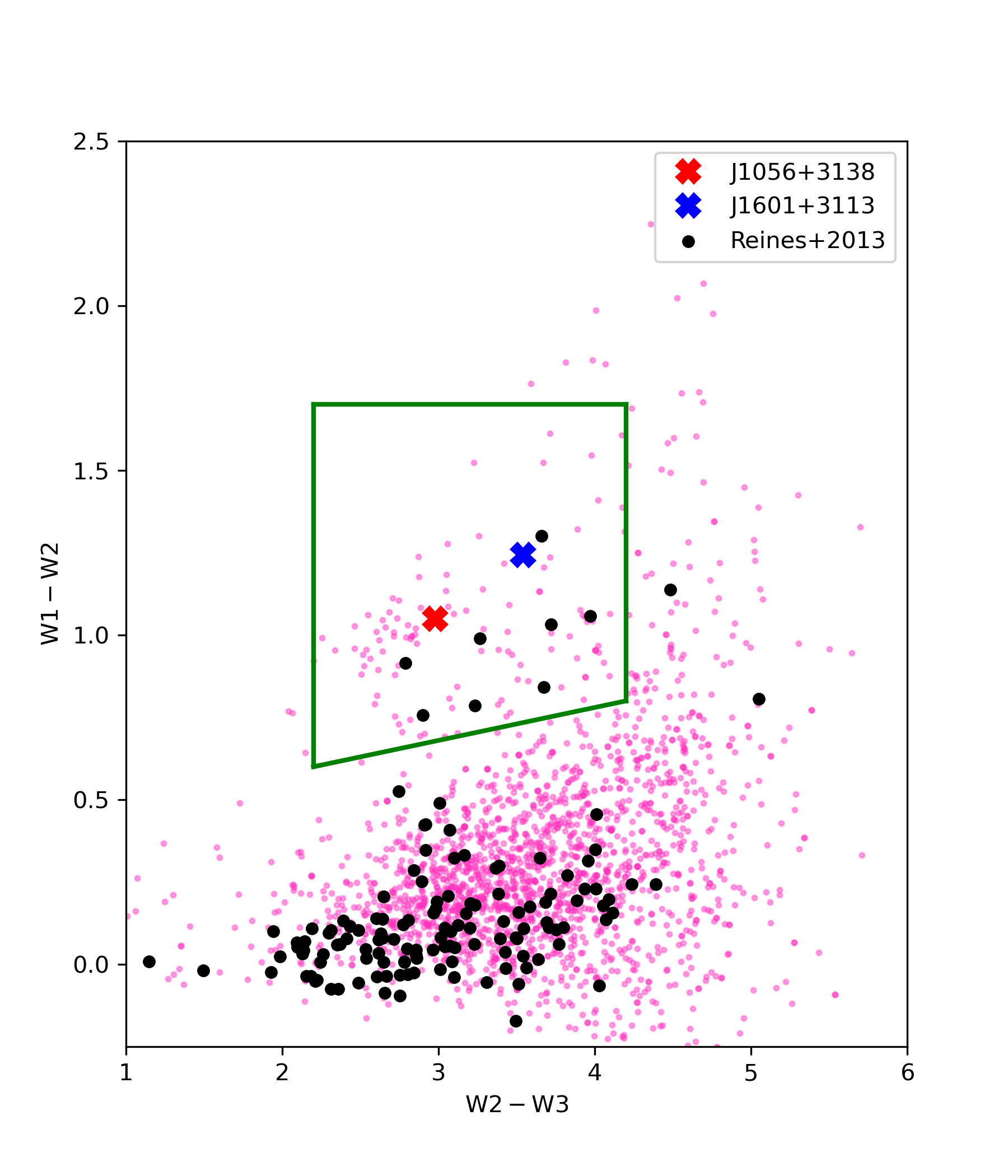}
\caption{Mid-infrared color-color diagram showing the placement of low metallicity galaxies (pink points) with comparable metallicity to J1601+3113 (blue `x') and J1056+3113 \citep[red `x';][]{cann2020}, as defined by log([\ion{N}{2}]/H$\alpha$) $< -1$, using emission line fluxes from the MPA/JHU catalog.  The green box outlines the \citet{jarrett2011} region occupied by dominant AGNs.  Targets from the \citet{reines2013} sample are in black, showing that our selection criteria probes a different subset of dwarf galaxies.}
\label{jarrett}
\end{figure}

While there is no evidence for an AGN in J1601+3113 based on its optical spectrum, its mid-infrared colors are suggestive of an AGN \citep{hainline2016}, using data from the Wide-field Infrared Survey Explorer (WISE) and the 3-band color cut from \citet{jarrett2011} (Figure \ref{jarrett}). While star formation in low metallicity galaxies can heat the dust to high temperatures \citep{griffith2011}, \citet{satyapal2018} show that it is very unlikely that a purely star forming low metallicity galaxy can meet the 3-band color cut, strongly suggesting the presence of an AGN in  J1601+3113. There are $\approx1500$ low metallicity galaxies that similarly show mid-infrared colors indicative of a dominant AGN, with masses up to three orders of magnitude less than the LMC, that are ideal candidates for follow-up observations. Approximately one-half of these are less massive than the lowest mass target in the \citet{reines2013} sample, so observations of this population could probe a new frontier of black hole mass. Note that mid-infrared color selection is biased towards the most dominant AGNs \citep{satyapal2021}, which comprise only a small fraction of the total AGN population \citep[see][]{yan2013}, so the discovery of an AGN in J1601+3113 may be the tip of the iceberg.

\subsection{Observations and Data Reduction}
Near-IR observations of J1601+3113 were obtained using the Gemini Near-InfraRed Spectrometer (GNIRS) using the cross dispersed mode with the 32 l/mm grating and nodding along a 0.45"x7" slit. The total integration time was 3600 seconds and obtained in clear conditions with 0.5" seeing. An A1V telluric star was observed at similar airmass.

The data reduction was carried out using the Gemini-provided IRAF package~\citep{cooke2005} for the flatfielding, sky subtraction, wavelength calibration, and spectral extraction. The IDL code, XTELLCOR\_GENERAL \citep{xtellcor}, was used for the telluric line removal and flux calibration. Line fluxes and uncertainties were determined from best-fit single Gaussian models to the emission lines, using a custom Python Bayesian maximum-likelihood code implemented in Python using the affine-invariant Markov Chain Monte Carlo (MCMC) package emcee \citep{emcee}. 

This object was also observed by SDSS on 12 May 2003 with the SDSS spectrograph. The optical data were reanalyzed to ensure accurate measurements of the BPT emission lines for BPT classification and to calculate upper limits of optical CLs, as well as to judge the presence of any broadened or outflow components to the emission lines. Spectral fitting was performed using version 7.1.1 of the open-source Python 3 code Bayesian AGN Decomposition Analysis for SDSS Spectra (BADASS) \citep{sexton2021}. In brief, BADASS implements emcee \citep{emcee} to obtain robust parameter fits and uncertainties, and utilizes a custom autocorrelation analysis to assess parameter convergence. The spectrum was run for a maximum of 25,000 MCMC iterations, with the mean of parameters converging around 20,000 iterations. The measured fluxes from BADASS for the emission lines reported in the MPA catalog are within good agreement.


\section{Results and Discussions}\label{sec:results}

\subsection{Near-Infrared Spectroscopy}
\begin{figure*}[h]
\centering
\includegraphics[width=\textwidth]{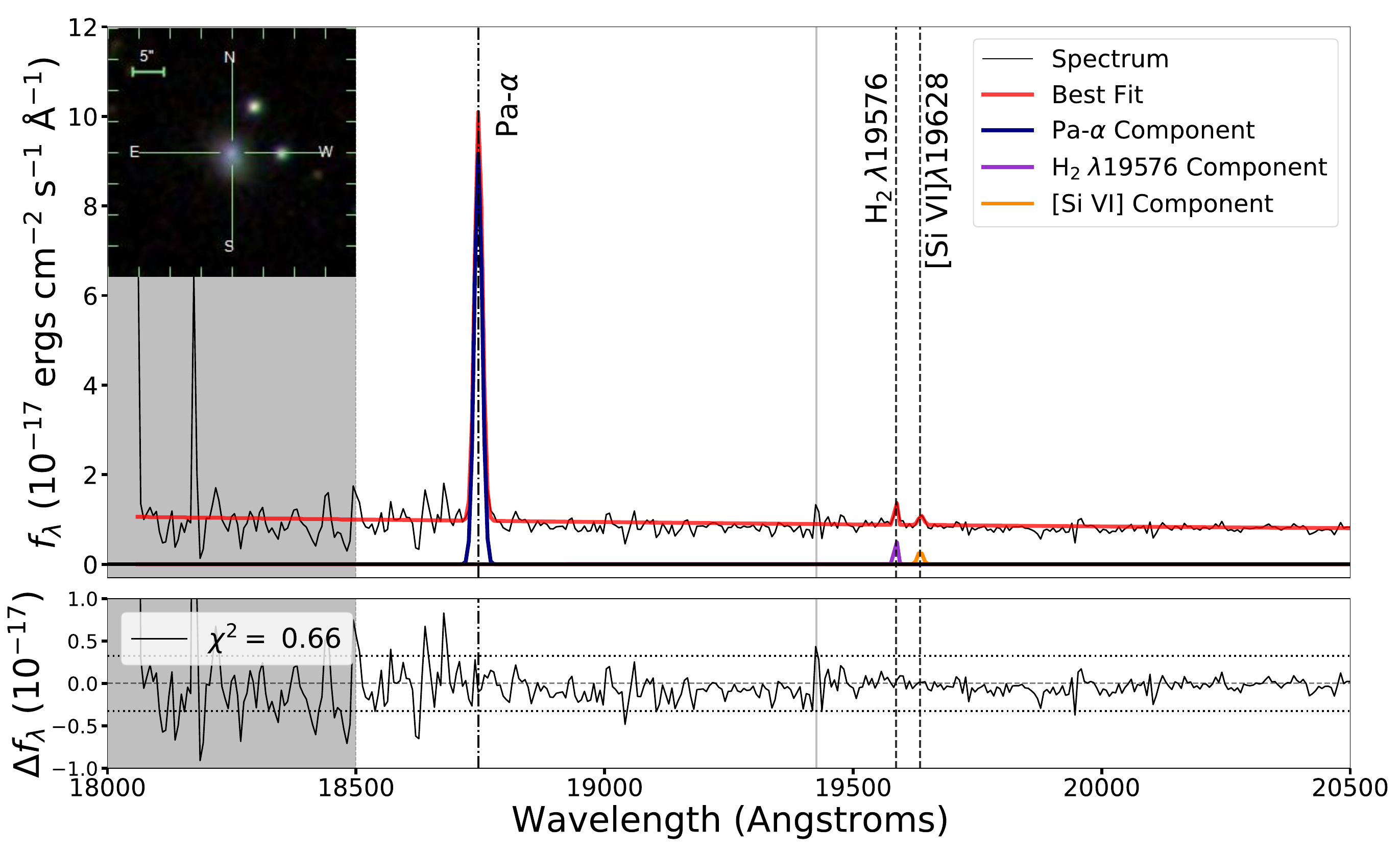}
\includegraphics[width=0.43\textwidth]{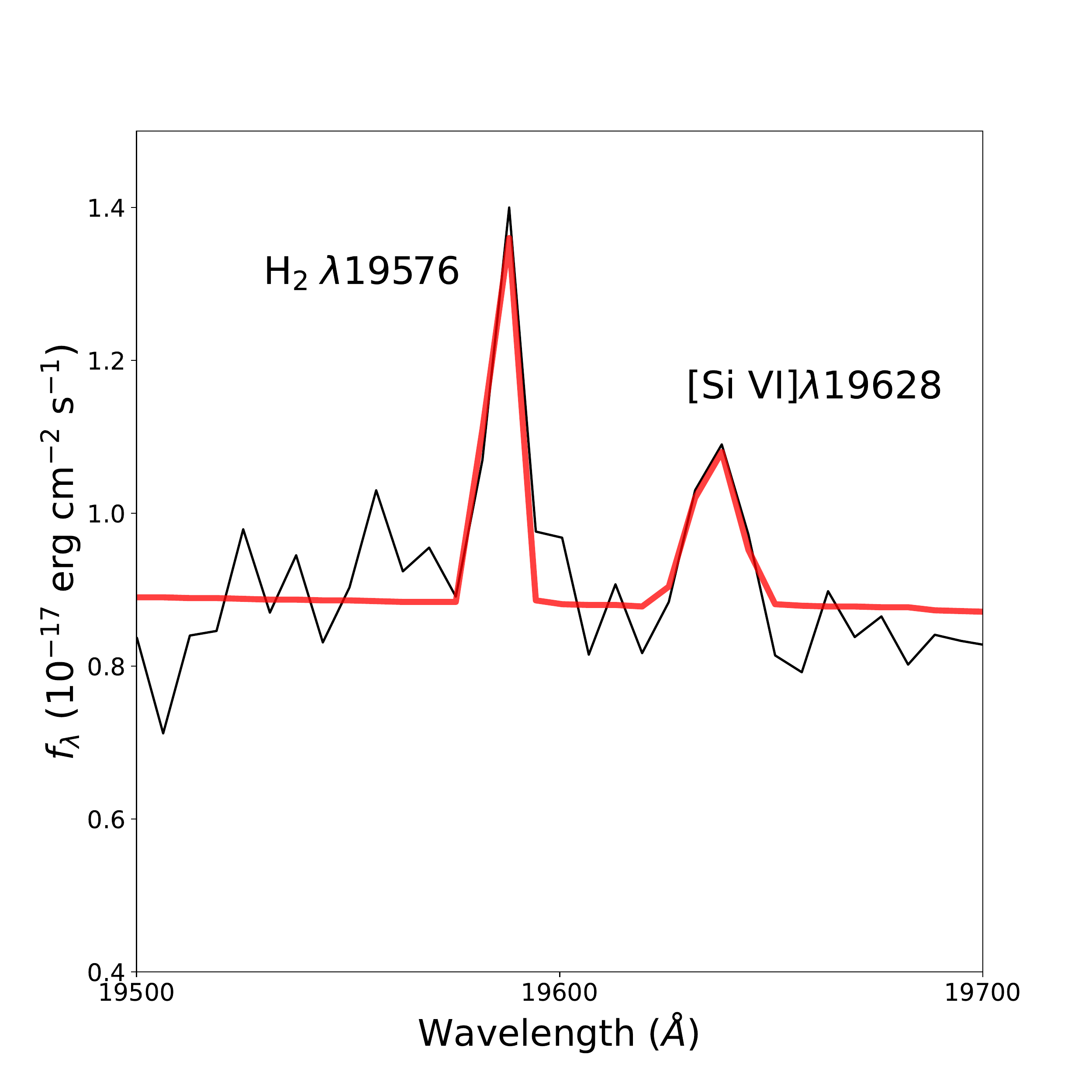} \includegraphics[width=0.55\textwidth]{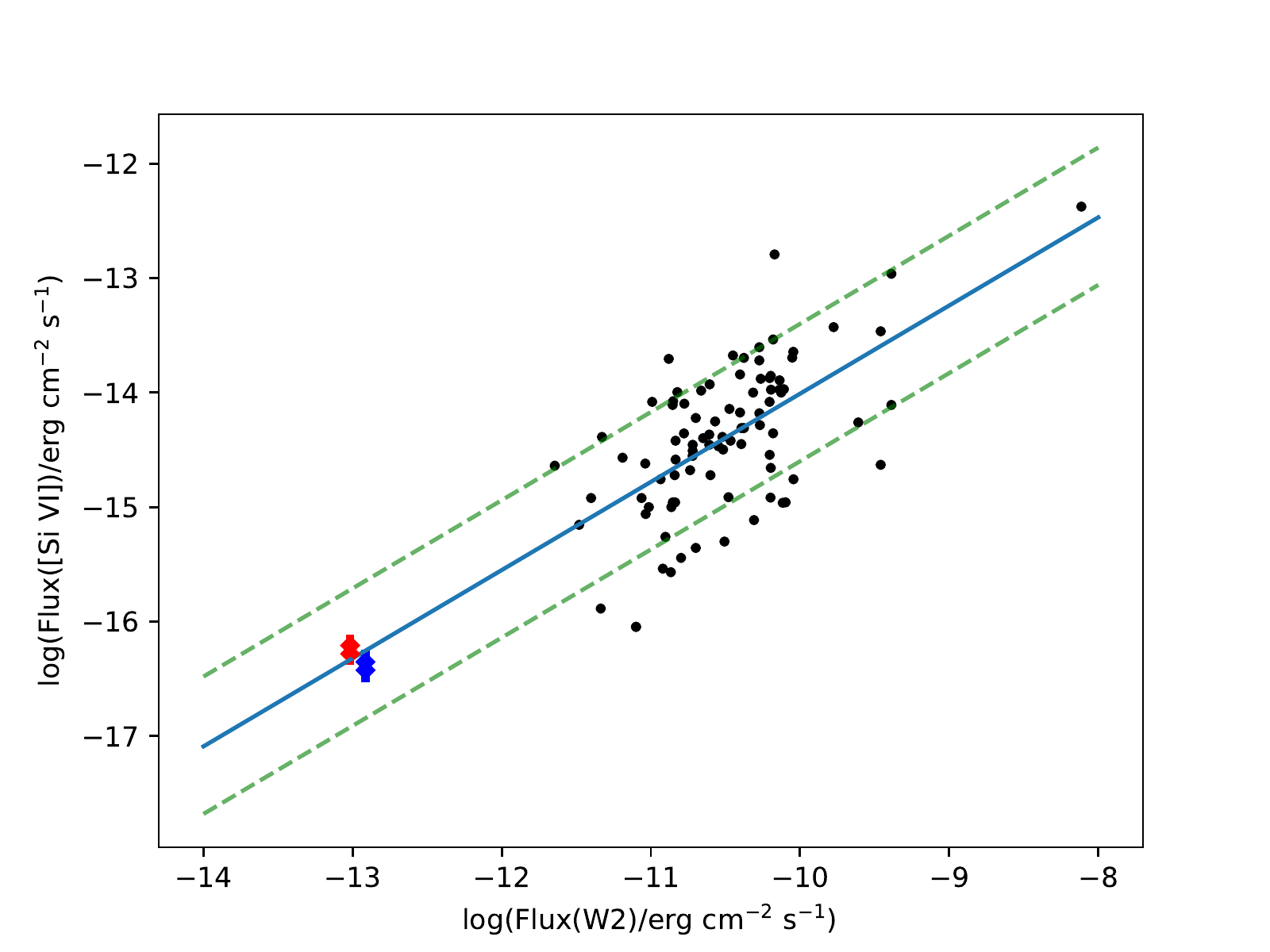}
\caption{\textit{Top:} The K-band spectra from Gemini \textit{GNIRS}.  Note the presence of a [\ion{Si}{6}] coronal line, as well as a Pa$\alpha$ and a H$_2$ line.  This is the first time a coronal line has been detected in a low metallicity dwarf galaxy with no evidence for an AGN in its optical spectrum. The grey shaded region corresponds to heavy telluric absorption, and the grey line denotes a poorly subtracted skyline. \textit{Bottom Left:} \textit{GNIRS} spectrum zoomed in around the [\ion{Si}{6}] line. \textit{Bottom Right:} The relation between W2 flux and [\ion{Si}{6}] flux, determined using the \citet{riffel2006,rodriguezardila2011,lamperti2017,mullersanchez2018} sample (black).  J1601+3113's (red 'x') fluxes are within the 1$\sigma$ scatter of the relation, as shown by the dotted green lines.}
\label{ir_spec}
\end{figure*}

The near-infrared K-band spectra is shown in Figure \ref{ir_spec}, and a list of observed fluxes and coronal line upper limits is given in Table \ref{fluxes}.  Most notably, we report a 3.6$\sigma$~[\ion{Si}{6}] coronal line with a flux of $4.0\pm1.1\times10^{-17}$~erg~cm$^{-2}$~s$^{-1}$.  The presence of this line, with an ionization potential of 167~eV is strong evidence for the presence of an AGN, as stellar sources do not produce enough high energy radiation to excite these ions \citep{satyapal2021}.  Further, the luminosity of this line is $9\times10^{37}$~erg~s$^{-1}$, comparable to the [\ion{Si}{6}] luminosities observed in near-infrared surveys of well-studied AGNs, ranging from $\approx10^{36}-10^{41}$~erg~s$^{-1}$ \citep[e.g.][]{lamperti2017,mullersanchez2018}. 
To further confirm the robustness of the detection, we looked at the relationship between [\ion{Si}{6}] detections and W2 fluxes in the literature \citep{riffel2006,rodriguezardila2011,lamperti2017,mullersanchez2018}, finding a correlation:
\begin{equation}\label{Eq 1}
    \log F_{\textrm{[\ion{Si}{6}]}} = (0.77 \pm 0.11)\times\log(F_{\textrm{W2}})-(6.31\pm0.55)
\end{equation}

Figure \ref{ir_spec} shows the relationship between the W2 flux and the [\ion{Si}{6}] flux for a  with J1601+3113 and J1056+3138 \citep{cann2020} overlaid.  As can be seen, J1601+3113 lies within the scatter of this relation, strongly suggesting that the coronal line flux is produced by an AGN.  This is the first time a coronal line has been detected in a low metallicity dwarf galaxy with no evidence for an AGN in its optical spectrum.  The widths of all NIR lines were fairly consistent with the optical lines ($\approx250$~km~s$^{-1}$), with some variation consistent with what has been observed in the literature, suggestive of differing gas dynamics between the various ISM phases \citep{rodriguezardila2005}.


\begin{table*}[t]
\caption{Near-infrared \textit{GNIRS} emission line fluxes}
\centering
\begin{tabular}{lcc}
\hline
\hline
\noalign{\smallskip}
    Line & Wavelength & Flux \\
    & $\mu$m & $10^{-17}~\mathrm{erg~cm^{-2}~s^{-1}}$\\
    \hline

    [\ion{S}{3}] & 0.9068 & $117.0 \pm 19.4$\\

    [\ion{S}{3}] & 0.9529 & $287.3 \pm 6.5$\\
    
    [\ion{S}{8}] & 0.9910 & $<40.4$\\
    
    Pa$\delta$ & 1.0047 & $18.7 \pm 6.4$\\
    
    [\ion{Fe}{13}] & 1.0747 & $<15.3$ \\
    
    \ion{He}{1} & 1.0830 & $125.1 \pm 5.2$\\
    Pa$\gamma$ & 1.0938 & $40.3 \pm 8.5$\\
    
    [\ion{S}{9}] & 1.2520 & $<33.6$ \\
    
    Pa$\beta$ & 1.2816 & $72.8 \pm 2.9$\\
    
    [\ion{Si}{10}] & 1.4300 & $<19.3$\\
    
    [\ion{Fe}{2}] & 1.6556 & $7.2 \pm 2.2$\\
    
    [\ion{Mg}{2}] & 1.6802 & $5.9 \pm 2.3$\\
    Pa$\alpha$ & 1.8756 & $182.4 \pm 2.2$\\
    H$_2$ & 1.9587 & $4.6 \pm 1.3$\\
    
    [\ion{Si}{11}] & 1.9320 & $<5.7$\\
    
    [\ion{Si}{6}] & 1.9628 & $4.0 \pm 1.1$\\
    
    [\ion{Al}{9}] & 2.0450 & $<5.3$\\
    
    \ion{He}{1} & 2.0576 & $10.1\pm0.8$\\
    
    Br$\gamma$ & 2.1650 & $15.0\pm1.8$\\
    
    [\ion{Ca}{8}] & 2.3210 & $<8.1$\\
    
    \noalign{\smallskip}
    \hline
    \noalign{\smallskip}       
     \end{tabular}
     \label{fluxes}
 \end{table*}

\subsection{Optical Spectroscopy}
The measured lines and upper limits from BADASS can be found in Table \ref{optical}.  There are no broad lines or outflow components detected, and there are no detections of optical CLs.   Given the [\ion{Si}{6}] flux, the non-detections are not unexpected given the typical ratios between [\ion{Si}{6}] and optical CLs typical of AGNs \citep[e.g.][]{oliva1994}.  For a more complete overview of the coronal line non-detections, see Section 3.6.


This, combined with its BPT star-forming designation, makes the AGN candidate in J1601+3113 invisible through traditional optical diagnostics. 

\begin{table*}[t]
\caption{Optical \textit{SDSS} emission line fluxes}
\centering
\begin{tabular}{lcc}
\hline
\hline
\noalign{\smallskip}
    Line  & Wavelength & Flux \\
    & \AA & $10^{-17}~\mathrm{erg~cm^{-2}~s^{-1}}$\\
    \hline
    [\ion{O}{2}] & 3726 & $527.0 \pm 11.4$\\
    
    H$\beta$ & 4861 & $619.7 \pm 6.5$\\
    
    [\ion{O}{3}] & 4959 & $760.5 \pm 2.1$\\
    
    [\ion{O}{3}] & 5007 & $2293.4 \pm 6.4$\\
    
    [\ion{Fe}{14}] & 5303 & $<13.3$\\
    
    [\ion{Fe}{7}] & 5722 & $<10.9$\\
    
    [\ion{Fe}{7}] & 6085 & $<8.7$\\
    
    [\ion{Fe}{10}] & 6374 & $<12.5$\\
    
    [\ion{N}{2}] & 6549 & $45.3 \pm 0.8$\\
    
    H$\alpha$ & 6563 & $1964.0 \pm 14.2$\\
    
    [\ion{N}{2}] & 6585 & $132.9 \pm 2.5$\\
    
    [\ion{S}{2}] & 6718 & $153.2 \pm 3.5$\\
    
    [\ion{S}{2}] & 6732 & $110.9 \pm 2.6$ \\
    
    
    \noalign{\smallskip}
    \hline
    \noalign{\smallskip}       
     \end{tabular}
     \label{optical}
 \end{table*}

\subsection{Extinction}

Dust extinction in J1601+3113's spectra was determined using the relative strengths of five observed hydrogen recombination lines from the \textit{GNIRS} spectrum.  Their fluxes and uncertainties can be found in Table \ref{fluxes}.  A Milky Way-like extinction curve ($R\mathrm{_{V}}$=3.1), implies minimal extinction ($A_V\approx-0.1$). 
The SDSS spectra shows an H$\beta$/H$\alpha$ ratio of $0.32\pm0.003$, implying negligible extinction.  The Pa$\alpha$/H$\alpha$ ratio of $0.09\pm0.001$
and an observed Pa$\alpha$/H$\beta$ ratio of $0.3\pm0.004$ \citep{hummerstory}, 
further confirm negligible obscuration.  

This implies that the lack of broad lines is likely due to AGN orientation, as opposed to host galaxy obscuration.  This result is comparable to \citet{lamperti2017}, in which only $\approx10\%$ of Seyfert 2s show near-infrared broad line emission.


\subsection{AGN Bolometric Luminosity and Mass Estimates}
Based on the BPT diagram the [\ion{O}{3}] flux, often used to estimate AGN bolometric luminosity,  was assumed to be of stellar origin. The [\ion{Si}{6}] flux was therefore used to estimate the bolometric luminosity by using a relationship with $L_{\mathrm{X,14-195~keV}}$ and in turn, $L_{\mathrm{X,14-195~keV}}$ vs $L_{\mathrm{bol}}$ \citep{winter2012,lamperti2017}.  This yields an $L_{\mathrm{X,14-195~keV}}\approx10^{41}$~erg~s$^{-1}$ and an $L_{\mathrm{bol}}\approx10^{42}$~erg~s$^{-1}$ with a scatter of 0.75 dex.  Note that the [\ion{O}{3}] emission predicted for an AGN of this bolometric luminosity is significantly lower than what is observed \citep{lamastra2009}, consistent with significant contamination from star formation to the [\ion{O}{3}] emission as expected based on its location on the BPT diagram.  The precise contribution however, is difficult to determine given that the published relations between the [\ion{O}{3}] and bolometric AGN luminosities are based on higher metallicity AGNs. A virial black hole mass cannot be determined owing to the lack of broad optical and near-IR emission lines.  Assuming the $M_*-M_{BH}$ relation of \citet{bohn2020} this yields $M_\mathrm{BH}\approx10^5$~M$_\odot$.  This implies a black hole accreting at $\approx10\%$ of the Eddington luminosity, though the relations above have been tested primarily for masses above $10^6$~M$_\odot$.




\subsection{Gas Metallicity Estimates}
As J1601+3113 is an optically normal, seemingly star-forming galaxy, that only shows its AGN nature in near- and mid-infrared diagnostics, the precise radiation field, and therefore, metallicity estimate holds a great deal of uncertainty, with estimates ranging from $\approx5-15$\% depending on the method used.  For the purposes of this study, an initial metallicity estimate was calculated using the same process as employed by \citet{izotov2007,izotov2008,cann2020}, using measurements of [\ion{O}{3}]5007, [\ion{O}{3}]4959, [\ion{O}{2}]3726 and H$\beta$.  Considering a Solar metallicity value of $12+\log(\mathrm{O}/\mathrm{H})$ of 8.69 \citep{asplund2006, groves2006}, we find a metallicity of $\approx10\%$ Solar for J1601+3113.  This is in agreement with our initial estimate based on its [\ion{N}{2}]/H$\alpha$ value, and is within the 1$\sigma$ scatter of the mass-metallicity relation \citep{yates2020}.

\subsection{Coronal Line Ratios}
The only coronal line detected in the optical and near-infrared spectra was the [Si~VI] line.  This is not unusual, since the lack of other coronal line detections is expected given the sensitivity of the SDSS and \textit{GNIRS} spectrum.  Indeed, in the hard X-ray selected AGN sample from \citet{lamperti2017}, [\ion{Si}{6}] is the most commonly detected near-infrared coronal line, detected over twice as often as the next most commonly detected line, [\ion{S}{9}].  Furthermore, when detected, the fluxes of other near-infrared CLs range up to an order of magnitude or more dimmer, well below the detection limits of \textit{GNIRS}.

Coronal line ratios, as discussed by \citet{cann2018}, may potentially provide insight to the black hole mass of an AGN.  This is because the lower mass black holes have hotter accretion disks, which in turn enhances the higher ionization line luminosities relative to the lower ionization lines.  We can investigate whether the optical and near-infrared spectra presented here can provide any information on the black hole mass in J1601+3113.  For example, the upper limit to the [\ion{Fe}{7}]~5722\AA/[\ion{Si}{6}] line flux ratio in J1601+3113 is 14.5. For comparison, this ratio ranges from $0.1 - 2.9$ in the subset of the hard X-ray selected AGN sample from \citet{lamperti2017} sample that has both near-infrared and previously published optical CLs. While this hard-X-ray selected sample has black hole masses that range from $10^{5.5}-10^8$~M$_\odot$, theoretical modeling with lower mass black holes suggests a decrease in the [\ion{Fe}{7}]~5722\AA/[\ion{Si}{6}] line flux ratio with black hole mass (Cann et al. 2021, in prep). The observed upper limit therefore is not sensitive enough to provide constraints on black hole mass from the [\ion{Fe}{7}]~5720\AA/[\ion{Si}{6}] line flux ratio.

Similarly, the [\ion{Si}{10}] line was not detected in our GNIRS spectrum. We obtain an upper limit on the [\ion{Si}{10}]/[\ion{Si}{6}] ratio of $\approx4.7$. Based on the compilation of near-infrared coronal line fluxes from the literature \citep{riffel2006,rodriguezardila2011,lamperti2017,mullersanchez2018}, this ratio ranges from $\approx0.3-4.8$ for black hole masses that range from $10^{6.6}-10^{8.2}$~M$_\odot$.  The upper limit on this line ratio for J1601+3113 is at the high end of this distribution, which again does not provide insight into the black hole mass.  The black hole mass derived in section 3.4 is consistent with this line ratio based on the models in \citet{cann2018}.
Likewise, the upper limit on the [\ion{Si}{11}]/[\ion{Si}{6}] ratio is $\approx1.4$.  Based on the compilation of near-infrared coronal line fluxes from the literature (\citep{lamperti2017}), this ratio ranges from $\approx0.1-0.7$ for black hole masses of $\approx10^7$~M$_\odot$.  While this upper limit is somewhat higher than those observed, the observed [\ion{Si}{11}]/[\ion{Si}{6}] is not as dramatically enhanced as predicted in the theoretical models presented in \citet{cann2018} for black holes substantially lower than $10^5$~M$_\odot$, suggesting that the optical and near-infrared spectrum of J1601+3113 is completely consistent with an AGN powered by a black hole with mass $\approx$ $10^5$~M$_\odot$, in line with expectations from its stellar mass.

\subsection{Alternative Scenarios}
While we have shown robust evidence for the presence of an accreting black hole in J1601+3113, here we outline in more detail alternative scenarios that we have ruled out.  

Optical CLs, such as [\ion{Ne}{5}], have been seen in extragalactic supernovae, and the [\ion{Si}{6}] coronal line has in fact been observed in non-AGN sources such as galactic supernovae and planetary nebulae, however the measured luminosities are low, ranging from $\approx10^{32}-10^{33}$ \citep[e.g.][]{benjamin1990}, at least five orders of magnitude below that observed in J1601+3113, and observationally undetectable outside our galaxy. Supernovae are typically accompanied by P Cygni line profiles and broad lines, while J1601+3113 does not show any sign of these features.  Further, enhanced [\ion{Fe}{2}] when compared to hydrogen recombination lines such as Pa$\beta$ is typically characteristic of shocks \citep{rodriguezardila2004}, however the ratio between [\ion{Fe}{2}]/Pa$\beta$ is $\approx0.09$ as seen in Table 1, which implies little contribution of the emission from shocks.  Using the [\ion{Fe}{2}]/Pa$\beta$ vs. H$_2$/Br$\gamma$ relation found in \citet{larkin1998,rodriguezardila2004,riffel2013}, J1601+3113 displays near-infrared line ratios indicative of a starburst galaxy, also indicating a lack of significant contribution to the emission by shocks. 

It has been theorized that non-AGN sources, such as Wolf-Rayet stars or shocks from starburst-driven winds \citep{schaerer1999,abel2008,allen2008}, can also be the source of CLs.  The observed line luminosities, however, are up to $4$ orders of magnitude weaker compared to optical [\ion{Ne}{5}] detections in AGNs \citep{izotov2012}.  Mid-infrared [\ion{Ne}{5}] can be up to $\approx8$ orders of magnitude dimmer in purely star-forming galaxies than in AGNs when compared to other common mid-infrared emission lines \citep{abel2008}. Moreover, [\ion{Ne}{5}] has a significantly lower ionization potential (99~eV) than [\ion{Si}{6}] (167~eV), further strengthening support for the presence of an AGN in J1601+3113.  Further, the CLs are typically accompanied by other signatures in the spectrum that point to the Wolf-Rayet stars, such as broadened [\ion{He}{2}]$\lambda4686$\AA \citep{izotov2007}, which are not present in the optical spectrum of J1601+3113. Also, there is a lack of the CN molecular absorption feature at 1.1$\mu$m, shown to serve as an indicator of younger to intermediate-aged stars. As the CN absorption feature is prominent in both Red Super Giants (0.01-0.03 Myr) and Asymptotic Giant Branch stars (0.3 – 1 Gyr), its absence implies a lack of evidence for recent star-forming activity that could produce shocks \citep{lancon2000,lancon2001,maraston2005,riffel2007}. Furthermore, the equivalent width of the Bracket gamma line is $\approx18$, which corresponds to a stellar population older than 5 Myr \citep{leitherer1999}. Finally, shocks are typically characterized by an enhancement of [\ion{Fe}{2}] emission in the near-infrared, and [\ion{O}{1}] and [\ion{S}{2}] in the optical, neither of which are seen in the spectra of J1601+3113.

In low metallicity galaxies, the stellar radiation field hardens with decreasing metallicity \citep{campbell1986}, potentially producing higher ionization lines than are typically found in pure star-forming galaxies.  This effect has been observed in several blue compact dwarf galaxies (BCDs), where CLs, such as optical [\ion{Ne}{5}], have been observed \citep{izotov2012}.  However, the CLs seen in these galaxies are much less luminous, and are of a lower ionization potential than the [\ion{Si}{6}] line (99~eV vs. 167~eV). Moreover, these galaxies are  characterized by a different region of MIR color-color space \citep[W1-W2=2.13-2.37, W2-W3=3.58-4.76;][]{izotov2011} compared to that of J1601+3113 seen in Figure \ref{jarrett}, believed to be due to a very recent starburst \citep{izotov2011}.

\subsection{Implications and Future Prospects}
The detection of an accreting SMBH in J1601+3113 has broad astrophysical implications for our understanding of the origins of SMBHs.  J1601+3113 is in one of the lowest metallicity galaxies known to host an AGN, and it is the first low metallicity dwarf galaxy to show a high ionization coronal line, despite showing no evidence for an AGN based on optical diagnostics.  This result suggests that the dearth of SMBHs in dwarf galaxies currently known may be more due to limitations in the currently used tools to find them rather than to an inherent lack of these objects in the Universe.

Low metallicity AGN (log([\ion{N}{2}]/H$\alpha$)$<-1.0$) are quite rare in current surveys.  \citet{groves2006} found only 40 low metallicity AGN candidates out of $\approx23,000$ Seyfert 2's from SDSS.  \citet{izotov2007,izotov2008,izotov2010} found broad line emission consistent with an AGN in several low metallicity galaxies, most of which were characterized as star-forming galaxies based on optical narrow line diagnostics. In a recent X-ray survey of low mass AGNs, only one was shown to be metal deficient \citep{schramm2013}.  \citet{cann2020} performed a multi-wavelength study on J1056+3138, a low metallicity AGN, that also shows the [\ion{Si}{6}] coronal line, as well as optical CLs, near-infrared and optical broad lines, and an X-ray point source. In a recent X-ray survey of dwarf galaxies,  61 showed X-ray point sources suggestive of low luminosity AGNs, 11 of which had [\ion{N}{2}]/H$\alpha$ ratios indicative of low metallicities \citep{birchall2020}, adding complementary support to a scenario in which AGNs do reside in low metallicity dwarf galaxies.

An order of magnitude lower mass than the LMC, J1601+3113 is remarkably low mass, as well as low metallicity.  In recent, large samples of optically selected AGNs in dwarf galaxies, only 2\%~\citep{reines2013} and 1\%~\citep{chilingarian2018} have masses equal to or lower than that of J1601+3113, making this target a particularly unique object. 

J1601+3113 is not identified as an AGN through optical broad line or narrow line diagnostics, such as the BPT diagram.  This is not unexpected given that many low metallicity galaxies have been shown to display optical line ratios indicative of star-forming galaxies \citep{cann2019}.  We note that optical variability has recently been used to search for AGNs in low mass galaxies \citep{baldassare2018,baldassare2020,martinez2020}. Long-lived stellar transients have been found to mimic AGNs in low metallicity dwarfs \citep{burke2020} producing broad lines that can persist over a decade, calling into question the diagnostic potential of optical variability in metal deficient dwarfs, though other signposts of stellar origin, such as P Cygni profiles, may show up in spectroscopic follow-up. Mid-infrared variability may hold some promise in this population \citep{secrest2020}.



Apart from the optical narrow line and broad line diagnostics, X-ray and radio techniques are also used to identify AGNs.  Radio observations in low metallicity star-forming dwarf galaxies, however, can be dominated by star-formation, hiding the AGN \citep{condon1991}.  Further, 80-85\% of AGNs of all masses do not show significant radio emission \citep{kellermann1989}, so the lack of radio emission does not necessarily imply a lack of AGN. 

AGN candidates are also often identified through X-ray diagnostics.  The predicted 2-10~keV X-ray luminosity of J1601+3113 is $\approx10^{41}$~erg~s$^{-1}$ based on its [\ion{Si}{6}] coronal line luminosity \citep{lamperti2017}, however there is $\approx0.5$ dex scatter in that relation. This luminosity is below the threshold of $10^{42}$~erg~s$^{-1}$ conventionally used to identify AGNs, showing that J1601+3113 would not be robustly identified through X-ray surveys.  
Furthermore, at the distance of J1601+3113, the predicted X-ray flux is below the sensitivity limit of \textit{Swift/BAT} and possibly even \textit{eRosita}.  This problem is exacerbated in low metallicity galaxies, where X-ray deficits have been observed \citep{simmonds2016,cann2020,burke2020}, where the X-ray luminosities may be up to 2 orders of magnitude less than that predicted for higher metallicity and more luminous AGNs, resulting in ambiguity with X-ray emission from stellar sources.

The results presented here prove that low mass AGNs can exist in galaxies that show no evidence for AGNs using traditional diagnostics, implying that the occupation fraction of AGNs in the low mass regime may be higher than inferred from current studies and highlighting a more promising method for uncovering them.  With the launch of JWST, infrared spectra with unprecedented sensitivity will be enabled, potentially uncovering a new population of low mass AGNs in low mass, low metallicity galaxies.



\acknowledgments

J.M.C. gratefully acknowledges support from an NSF GRFP.   G.C. and T.B. acknowledge partial support for this project provided by the National Science Foundation, under grant No. AST 1817233. L.B. acknowledges support from NSF grant AST-1715413. The  authors  would  like  to  thank  V.  Ma\v{c}ka  and  R.T. Gatto for their insightful discussions and support in the analysis of the work presented.  
The authors would also like to thank the anonymous referee for their insightful comments and suggestions.

This work is based on observations obtained through Program ID GN-2020A-Q233 at the international Gemini Observatory, a program of NSF’s NOIRLab, which is managed by the Association of Universities for Research in Astronomy (AURA) under a cooperative agreement with the National Science Foundation on behalf of the Gemini Observatory partnership: the National Science Foundation (United States), National Research Council (Canada), Agencia Nacional de Investigaci\'{o}n y Desarrollo (Chile), Ministerio de Ciencia, Tecnolog\'{i}a e Innovaci\'{o}n (Argentina), Minist\'{e}rio da Ci\^{e}ncia, Tecnologia, Inova\c{c}\~{o}es e Comunica\c{c}\~{o}es (Brazil), and Korea Astronomy and Space Science Institute (Republic of Korea). This data was processed using the Gemini IRAF package.

This work was enabled by observations made from the Gemini North telescope, located within the Maunakea Science Reserve and adjacent to the summit of Maunakea. We are grateful for the privilege of observing the Universe from a place that is unique in both its astronomical quality and its cultural significance.  The authors wish to recognize and acknowledge the very significant cultural role and reverence that the summit of Mauna Kea has always had within the indigenous Hawaiian community. We are most fortunate to have the opportunity to conduct observations from this mountain.

\vspace{5mm}
\facilities{Gemini, SDSS}

\end{document}